# CORDIC-based Architecture for Powering Computation in Fixed Point Arithmetic

Nia Simmonds[†], Joshua Mack[*]
[†]Dept. of Electrical Engineering and Computer Science
Case Western Reserve University, OH, USA
[†]Electrical and Computer Engineering Department
The University of Arizona, Tucson, AZ, USA
nia.simmonds@case.edu, jmack2545@email.arizona.edu

Sam Bellestri[+], Daniel Llamocca[★]
Electrical and Computer Engineering Department
[+]The University of Alabama, AL, USA
[★]Oakland University, Rochester, MI, USA
sdbellestri@crimson.ua.edu, llamocca@oakland.edu

*Abstract*—We present a fixed point architecture (source VHDL code is provided) for powering computation. The fully customized architecture, based on the expanded hyperbolic CORDIC algorithm, allows for design space exploration to establish trade-offs among design parameters (numerical format, number of iterations), execution time, resource usage and accuracy. We also generate Pareto-optimal realizations in the resource-accuracy space: this approach can produce optimal hardware realizations that simultaneously satisfy resource and accuracy requirements.

## I. INTRODUCTION

The powering function frequently appears in many scientific and engineering applications. Accurate software routines are readily available, but they are often too slow for real-time applications. On the other hand, dedicated hardware implementations using fixed-point arithmetic are attractive as they can exhibit high performance, and low resource usage.

Direct computation of $x^y$ in hardware usually requires table lookup methods and polynomial approximations, which are not efficient in terms of resource usage. The authors in [1] propose an efficient composite algorithm for floating point arithmetic. The work in [2] describes the implementation of $x^y = e^{y \ln x}$ in floating point arithmetic using a table-based reduction with polynomial approximation for $e^x$ and a range reduction method for $\ln x$. Alternatively, we can efficiently implement $e^x$ and $\ln x$ using the well-known hyperbolic CORDIC algorithm [3]: a fixed point architecture with an expanded range of convergence is presented in [4], and a scale-free fixed point hardware is described in [5]. There are other algorithms that outperform CORDIC under certain conditions: the BKM algorithm [6] generalizes CORDIC and features some advantages in the residue number system, and a modification of CORDIC is proposed in [7] for faster computation of $e^x$. All the listed methods impose a constraint on the domain of the functions.

This work presents a fixed-point hardware for $x^y$ computation. We use the hyperbolic CORDIC algorithm with expanded range of convergence [8] to first implement $e^x$ and $\ln x$, and then $x^y = e^{y \ln x}$. Compared to a floating point implementation presented in [9], a fixed point hardware reduces resource usage at the expense of reduced dynamic range and accuracy. The main contributions of this work include:

- *Open-source, generic and customized architecture validated on an FPGA:* The architecture, developed at the register transfer level in fully parameterized VHDL code, can be implemented on any existing hardware technology (e.g.: FPGA, ASIC, Programmable SoC).
- *Design space exploration:* We explore trade-offs among resources, performance, accuracy, and hardware design parameters. In particular, we study the effect of the fixed point format on accuracy.
- *Pareto-optimal realizations based on accuracy and resource usage*: By generating the set of optimal (in the multi-objective sense) architectures, we can optimally manage resources by modifying accuracy requirements.

The paper is organized as follows: Section II details the CORDIC-based computation of $x^y$. Section III describes the architecture for $e^x$, $\ln x$, and $x^y$. Section IV details the experimental setup. Section V presents results in terms of execution time, accuracy, resources, and Pareto-optimal realizations. Conclusions are provided in Section VI.

## II. CORDIC-BASED CALCULATION OF $x^y$

Here, we explain how the expanded hyperbolic CORDIC algorithm is used to compute $\ln x$ and $e^x$, from which we get $x^y = e^{y \ln x}$. We then analyze the input domain bounds of $x^y$ resulting from the expanded hyperbolic CORDIC algorithm.

### A. Expanded hyperbolic CORDIC to compute $\ln x$ and $e^x$

The original hyperbolic CORDIC algorithm has a limited range of convergence. To address this issue, the expanded hyperbolic CORDIC algorithm [8] introduces additional iterations with negative indices. The value of $\delta_i$ depends on the operation mode:

$$i \leq 0: \begin{cases} x_{i+1} = x_i + \delta_i y_i(1 - 2^{i-2}) \\ y_{i+1} = y_i + \delta_i x_i(1 - 2^{i-2}) \\ z_{i+1} = z_i - \delta_i \theta_i, \theta_i = Tanh^{-1}(1 - 2^{i-2}) \end{cases} \quad (1)$$

$$i > 0: \begin{cases} x_{i+1} = x_i + \delta_i y_i 2^{-i} \\ y_{i+1} = y_i + \delta_i x_i 2^{-i} \\ z_{i+1} = z_i - \delta_i \theta_i, \theta_i = Tanh^{-1}(2^{-i}) \end{cases} \quad (2)$$

$Rotation: \delta_i = -1 \text{ if } z_i < 0; +1, otherwise$
$Vectoring: \delta_i = -1 \text{ if } x_i y_i \geq 0; +1, otherwise$ $\quad (3)$

This material is based upon work supported by the National Science Foundation under Grant No. EEC-1263133.

TABLE I. CORDIC CONVERGENCE BOUNDS FOR THE ARGUMENT OF THE FUNCTIONS AS M INCREASES. THE ORIGINAL CORDIC CASE IS INCLUDED.

| $M$ | $\cosh x, \sinh x, e^x$ | $\ln x$ |
|---|---|---|
| Original CORDIC | $[-1.11820, 1.11820]$ | $(0, 9.35958]$ |
| 0 | $[-2.09113, 2.09113]$ | $(0, 65.51375]$ |
| 1 | $[-3.44515, 3.44515]$ | $(0, 982.69618]$ |
| 2 | $[-5.16215, 5,16215]$ | $(0, 3.04640 \times 10^4]$ |
| 3 | $[-7.23371, 7.23371]$ | $(0, 1.91920 \times 10^6]$ |
| 4 | $[-9.65581, 9.65581]$ | $(0, 2.43742 \times 10^8]$ |
| 5 | $[-12.42644, 12.42644]$ | $(0, 6.21539 \times 10^{10}]$ |
| 6 | $[-15.54462, 15,54462]$ | $(0, 3.17604 \times 10^{13}]$ |
| 7 | $[-19.00987, 19.00987]$ | $(0, 3.24910 \times 10^{16}]$ |
| 8 | $[-22.82194, 22.82194]$ | $(0, 6.65097 \times 10^{19}]$ |
| 9 | $[-26.98070, 26,98070]$ | $(0, 2.72357 \times 10^{23}]$ |
| 10 | $[-31.48609, 31.48609]$ | $(0, 2.23085 \times 10^{27}]$ |

There are $M + 1$ negative iterations ($i = -M, ..., -1, 0$) and $N$ positive iterations ($i = 1, 2, ..., N$). The iterations $4, 13, 40, ..., k, 3k + 1$ must be repeated to guarantee convergence. For sufficiently large $N$, the values of $x_n, y_n, z_n$ converge to:

$$Rotation: \begin{cases} x_n = A_n(x_{in}\cosh z_{in} + y_{in}\sinh z_{in}) \\ y_n = A_n(x_{in}\cosh z_{in} + y_{in}\sinh z_{in}) \\ z_n = 0 \end{cases} \quad (4)$$

$$Vectoring: \begin{cases} x_n = A_n\sqrt{x_{in}^2 - y_{in}^2} \\ y_n = 0 \\ z_n = z_{in} + \tanh^{-1}(y_{in}/x_{in}) \end{cases} \quad (5)$$

$$A_n = \left(\prod_{i=-M}^{0} \sqrt{1 - (1 - 2^{i-2})^2}\right) \prod_{i=1}^{N} \sqrt{1 - 2^{-2i}} \quad (6)$$

Note that $x_{in} = x_{-M}, y_{in} = y_{-M}, z_{in} = z_{-M}$. To get $e^\alpha = \cosh\alpha + \sinh\alpha$, we use $x_{in} = y_{in} = 1/A_n$, $z_{in} = \alpha$ in the rotation mode. To get $(\ln\alpha)/2 = \tanh^{-1}(\alpha - 1/\alpha + 1)$, we use $x_{in} = \alpha + 1$, $y_{in} = \alpha - 1$, $z_{in} = 0$ in the vectoring mode.

The parameter $M$ controls the range of convergence of the expanded hyperbolic CORDIC: $[-\theta_{max}(M), \theta_{max}(M)]$. This is the bound on the domain of $\cosh/\sinh/e^x$ and the range of $\tanh^{-1}$. For $\ln x$, the argument is bounded by $\left(0, e^{\theta_{max}(M) \times 2}\right]$. Table I shows, as $M$ increases, how the expanded CORDIC dramatically expands the argument bounds of $e^x$ and $\ln x$ The expanded CORDIC is thus crucial for proper $x^y$ computation.

*B. Computation of $x^y$*

To compute $x^y = e^{y \ln x}$, we first use CORDIC in the vectoring mode with $x_{in} = x + 1$, $y_{in} = x - 1$, $z_{in} = 0$ to get $z_n = (\ln x)/2$. We then apply $z_n \times 2y = y \ln x$. Finally, we use CORDIC in the rotation mode with $x_{in} = y_{in} = 1/A_n$, $z_{in} = y \ln x$ to get $x_n = e^{y \ln x} = x^y$.

Fig. 1 depicts the input domain bound (area bounded by the curve) as a function of $M$ for $x^y = e^{y \ln x}$, which is given by $|y \ln x| \leq \theta_{max}(M)$. These are the $(x, y)$ values for which $x^y$ converges. Note the asymptotes when $x \to 1$ (as $\ln x \to 0$) and $y \to 0$. The input domain does not include $x \leq 0$, as $\ln x$ is undefined for $x \leq 0$. Thus, the algorithm can only compute $x^y$ for $x > 0$. For $x < 0$ and an integer $y$, we can compute $|x|^y$ and $(-1)^y$ separately; for non-integer $y$, we cannot compute $x^y$ as the result is a complex number.

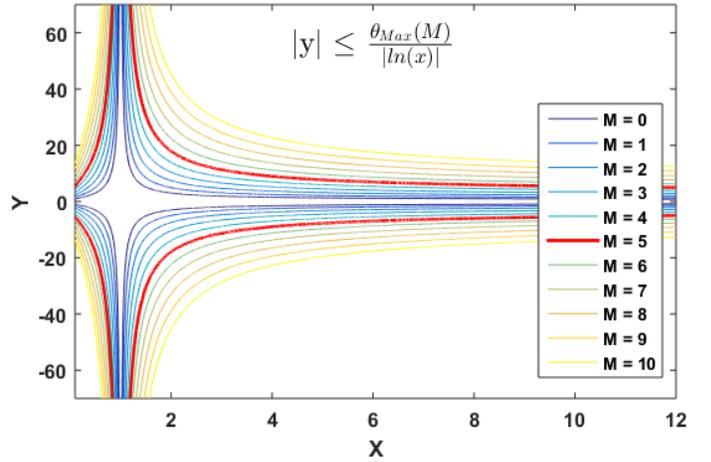

Figure 1. Domain of convergence for $x^y$ as a function of $M$: the domain grows as $M$ increases. The case $M = 5$ is highlighted.

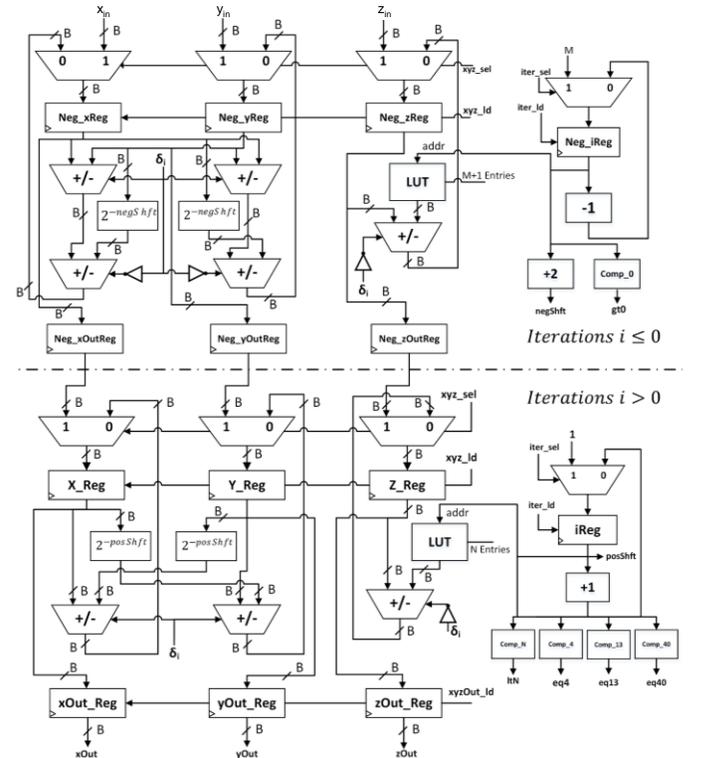

Figure 2. Parameterized expanded hyperbolic CORDIC engine. $B$: bit-width. $N$: number of positive iterations, $M + 1$: number of negative iterations.

### III. FIXED-POINT ITERATIVE ARCHITECTURE FOR $x^y$

Here, we describe the fixed-point hardware that computes $x^y = e^{y \ln x}$. This architecture is based on an expanded hyperbolic CORDIC engine that can compute $e^x$ and $\ln x$.

*A. Hyperbolic CORDIC engine*

Fig. 2 depicts the hyperbolic CORDIC engine. The top stage implements the $M + 1$ negative iterations, while the bottom stage implements the $N$ positive iterations. By proper selection of $x_{in}, y_{in}, z_{in}$ and the operation mode, this architecture can compute various functions (e.g.: $e^x, \cosh, \sinh, atanh$).

Each stage utilizes two barrel shifters, a look-up table (LUT), and multiplexers. The top stage requires five adders, while the bottom one requires three adders. A state machine controls the iteration counter for $i$, the add/sub input of the adders, the loading of the registers, and the multiplexer selectors.

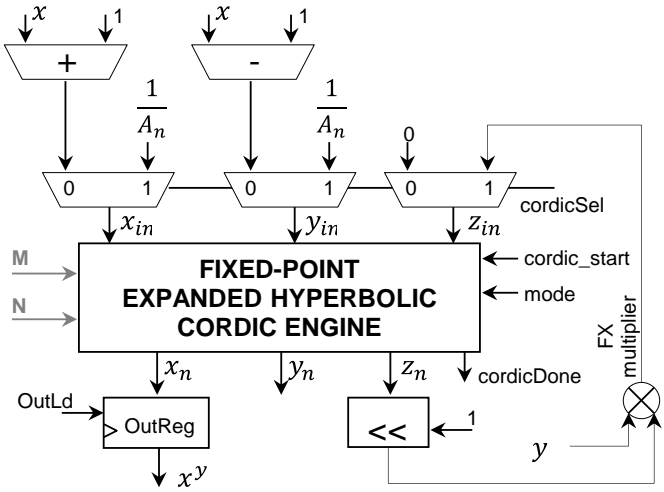

Figure 3. Block Diagram for Powering ($x^y$) computation. The expanded hyperbolic CORDIC engine is utilized twice.

We use the fixed point format $[B\ FW]$ through all the datapath, with $IW = B - FW$ integer bits and $FW$ fractional bits. The customized hyperbolic CORDIC architecture allows the user to modify the design parameters: number of bits ($B$), number of fractional bits ($FW$), number of positive iterations ($N$), and number of negative iterations ($M + 1$). The use of fixed point arithmetic optimizes resource usage. However, as it features a small numeric range, we might not be able to use the entire input domain of the algorithm (see Table I).

### B. Architecture for Powering Computation: $x^y$

Fig. 3 depicts the block diagram of the circuit that implements $x^y = e^{y \ln x}$. Note that the same bit-width ($B$) is used throughout the architecture. This circuit utilizes one hyperbolic CORDIC engine in two steps:

First, we load $x_{in} = x + 1$, $y_{in} = x - 1$, $z_{in} = 0$ onto the CORDIC engine in the vectoring mode, so that $z_n = \ln x / 2$. To get $x + 1$, $x - 1$, we use adders with a constant input. A shifter generates $\ln x$. A fixed point multiplier then computes $y \ln x$, which is fed back into the CORDIC engine. In the second step, we load $x_{in} = y_{in} = 1/A_n$, $z_{in} = y \ln x$ onto the CORDIC engine in the rotation mode, so that we get $x_n = e^{y \ln x} = x^y$.

The design parameters of the $x^y$ architecture are: bit-width ($B$), fractional bit-width ($FW$), number of positive iterations ($N$), and number of negative iterations ($M + 1$). This allows for fine control of accuracy, execution time, and resources.

## IV. EXPERIMENTAL SETUP

### A. Selection of parameters for design space exploration

The parameterized VHDL code allows for the generation of a space of hardware profiles by varying the design parameters. We consider: $B$ (24, 28, 32, 36, 40, 48, 52, 56, 60, 64, 68, 72, 76) and $N$ (8, 12, 16, 20, 24, 28, 32, 36, 40). A discussion on the format $[B\ FW]$ is presented in Section IV.C. For simplicity's sake, we fix $M = 5$ (6 negative iterations). Each of the functions $e^x, \ln x, x^y$ requires a different architecture. For each function, we generate $13 \times 9 = 117$ different hardware profiles. Results are obtained for every hardware profile and for every function.

### B. Generation of input signals

For $e^x$ and $\ln x$, we selected 1000 equally spaced points in the allowable input domain listed in Table I for $M = 5$.

For $x^y$ testing, we used $150 \times 10$ linearly spaced $(x, y)$ points, where $x \in [e^{-\theta_{max}(M=5)}, e^{\theta_{max}(M=5)}]$ (allowed $x$ interval when $|y| = 1$). The interval for $y$ varies according to $x$ as per the formula $|y \ln x| \leq \theta_{max}(M = 5)$.

### C. Selection of fixed point formats

By selecting the integer bit-width ($IW$) and the fractional bit-width ($FW$), a custom fixed point format $[B\ FW]$ is defined, where $B = FW + IW$. The range of values is given by $[-2^{B-FW-1}, 2^{B-FW-1} - 2^{-FW}]$. Table II list the formats we selected for our experiments ($B = 24 \rightarrow 76$, in increments of 4) along with the maximum value ($2^{B-FW-1} - 2^{-FW}$), the resolution ($2^{-FW}$), and the dynamic range ($2^{B-1}$) in dB.

For proper fixed point representation of the input, intermediate, and output values of $e^x, \ln x, x^y$, a large number of bits is required. For the selected input domains of Section IV.B, the $e^x$ and $x^y$ functions required 20 integer bits, while the $\ln x$ function required 37 integer bits. As for the number of fractional bits, we start with 8 bits and then keep increasing it by 4 up to a maximum of 32 bits.

For $e^x$, Fig. 4 plots $x_i$ for each iteration ($i = -5 \rightarrow 40$) for various $x$ values. Note that $x_i$ can be as twice as the final value $x_N$. The largest $e^{x=\theta_{max}(M=5)}$ needs $IW = 19$, thus we need $IW = 20$ to properly represent the intermediate values. To assess the loss in accuracy, we included a format with $IW < 20$.

For $\ln x$, $IW = 37$ bits are required to cover the input domain. Thus, we included cases with $B > 68$ bits in Table II.

The scaling factor provided as a constant input to the architecture, depends on $N$ and $M$ as per (6). The VHDL code was synthesized on a Xilinx® Zynq-7000 XC7Z010 SoC (ARM processor + FPGA fabric) that runs at 125MHz.

TABLE II. LIST OF FIXED POINT FORMATS USED IN OUR EXPERIMENTS
IW: INTEGER WIDTH, FW: FRACTION WIDTH. $B = FW + IW$

| [B FW] | IW | Maximum value | Resolution | Dyn. Range |
|---|---|---|---|---|
| [24 8] | 16 | $3.277 \times 10^4$ | $3.906 \times 10^{-3}$ | 138.5 dB |
| [28 8] | 20 | $5.243 \times 10^5$ | $3.906 \times 10^{-3}$ | 162.6 dB |
| [32 12] | 20 | $5.243 \times 10^5$ | $2.441 \times 10^{-4}$ | 186.6 dB |
| [36 16] | 20 | $5.243 \times 10^5$ | $1.526 \times 10^{-5}$ | 210.7 dB |
| [40 20] | 20 | $5.243 \times 10^5$ | $9.536 \times 10^{-7}$ | 234.8 dB |
| [44 24] | 20 | $5.243 \times 10^5$ | $5.960 \times 10^{-8}$ | 258.9 dB |
| [48 28] | 20 | $5.243 \times 10^5$ | $3.725 \times 10^{-9}$ | 283.0 dB |
| [52 32] | 20 | $5.243 \times 10^5$ | $2.328 \times 10^{-10}$ | 307.1 dB |
| [56 32] | 24 | $8.388 \times 10^6$ | $2.328 \times 10^{-10}$ | 331.1 dB |
| [60 32] | 28 | $1.342 \times 10^8$ | $2.328 \times 10^{-10}$ | 355.2 dB |
| [64 32] | 32 | $2.147 \times 10^9$ | $2.328 \times 10^{-10}$ | 379.3 dB |
| [68 32] | 36 | $3.436 \times 10^{10}$ | $2.328 \times 10^{-10}$ | 403.4 dB |
| [72 32] | 40 | $5.497 \times 10^{11}$ | $2.328 \times 10^{-10}$ | 427.5 dB |
| [76 32] | 44 | $8.796 \times 10^{12}$ | $2.328 \times 10^{-10}$ | 451.5 dB |

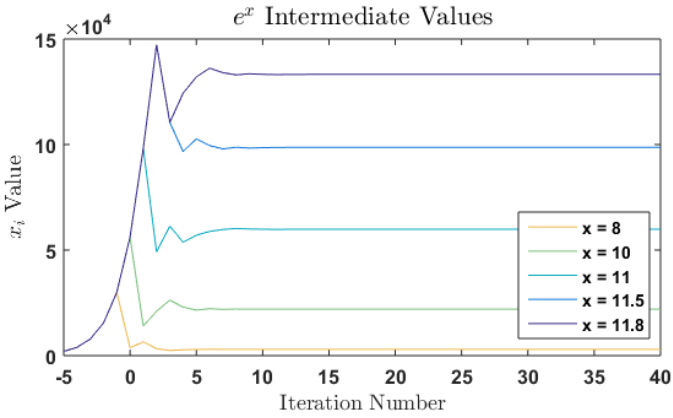

Figure 4. $e^x$ CORDIC computation: $x_i$ vs. the iteration number for $x = 8, 10, 11, 11.8$. Note that the intermediate values can be larger than the result $x_N$.

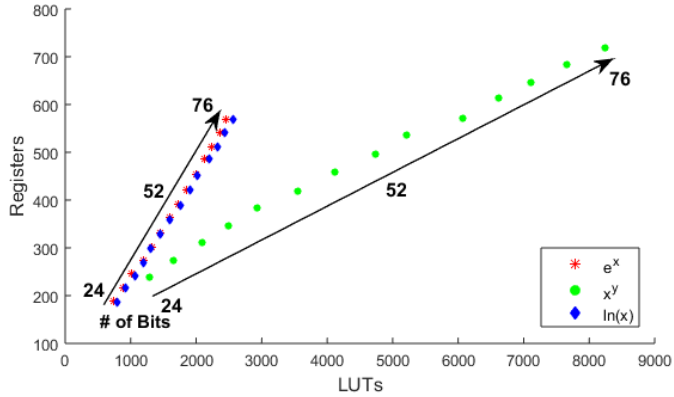

Figure 5. Resource utilization for $e^x$, $\ln x$, $x^y$ functions. Device: Zynq-7000 XC7Z010 SoC with 35,200 registers and 17,600 6-input LUTs

## V. RESULTS AND ANALYSIS

### A. Execution Time

For $e^x$ and $\ln x$, we require one cycle per iteration. The output register of each stage requires two extra cycles. For $x^y$, we add the execution times for $e^x$ and $\ln x$, and an extra cycle to place the final result on the output register. Execution time (in number of cycles) depends on $N$ and $M$, and it is given by:

$$Exec.time\ (e^x, \ln x) = M + 1 + N + v(N) + 2 \quad (7)$$

$$Exec.time\ (x^y) = 2(M+1) + 2N + 2 \times v(N) + 5 \quad (8)$$

$v(N)$ refers to the number of repeated iterations (see Section II). Table III lists execution time (ns) for a clock frequency of 125 MHz for different values of $N$ for the given functions.

TABLE III. EXECUTION TIME (NS) FOR $e^x$, $\ln x$, $x^y$. FREQUENCY: 125 MHZ.

| Function | N (number of positive iterations), M=5 | | | | | | | |
|---|---|---|---|---|---|---|---|---|
| | 8 | 12 | 16 | 20 | 24 | 32 | 36 | 40 |
| $e^x / \ln x$ | 136 | 168 | 208 | 240 | 272 | 336 | 368 | 408 |
| $x^y$ | 280 | 344 | 424 | 488 | 552 | 680 | 744 | 824 |

### B. Resource usage

Fig. 5 shows resource usage only in terms of 6-input LUTs and 1-bit registers for the fourteen bit-widths of Table II and for the $e^x$, $\ln x$, and $x^y$ architectures. As the number of bits grow, so does the resources. The LUT increase is more pronounced, indicating a large combinational cost. Here, we fixed $M = 5$.

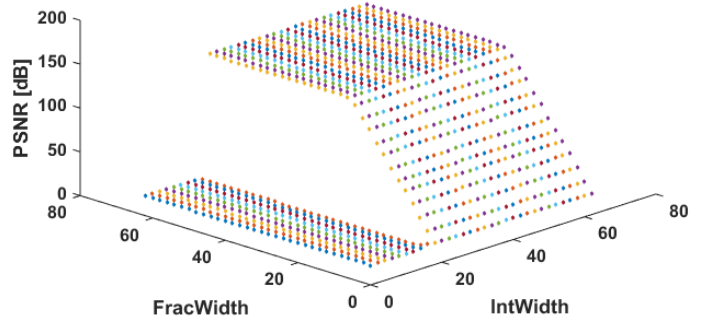

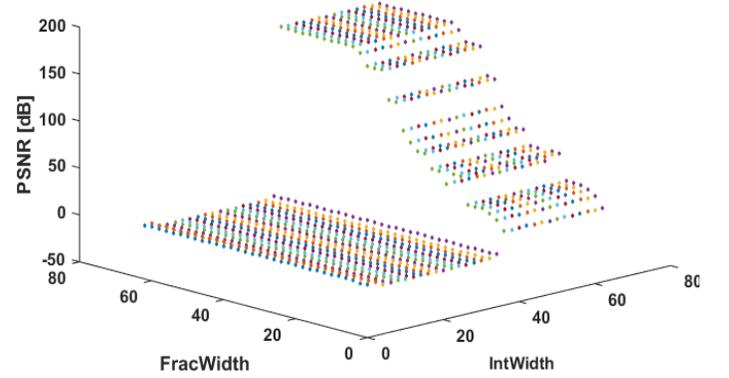

Figure 6. Accuracy results vs. the number of integer bits ($IW$) and fractional bits ($FW$) for the $e^x$ and $\ln x$ fixed point architectures. $N = 40, M = 5$.

The effect of $N$ on resource usage is negligible: $N$ only affects the size of the LUT for the angles and the state machine.

### C. Accuracy

For accuracy, we use the peak signal-to-noise ratio: $PSNR(dB) = 10 \times \log_{10}(maxval^2/MSE)$, where MSE is the mean squared error between the results of our architecture and the reference results provided by the MATLAB® built-in function in double floating point precision. $maxval$ is defined as the largest value of the fixed point output format. However, for consistency, we use the shortest fixed point format that can represent the largest output value for each function (this might differ from the one in Table II).

To validate our selection of fixed point formats, Fig. 6 shows accuracy results for $N = 40$ for $e^x$ and $\ln x$ in the input domain of Section IV.B. Note the very poor accuracy when $IW < 20$ and $IW < 37$ for $e^x$ and $\ln x$ respectively. We can also see the effect of the number of fractional bits on accuracy.

Figs. 7, 8, and 9 plot accuracy as a function of the number of positive iterations ($N$) and the bit-width ($B$) for $e^x$, $\ln x$, and $x^y$ respectively. In each case, note how the PSNR values stabilize after a certain number of iterations. For $e^x$, the case $B = 24$ yields poor results regardless of the value of $N$. For $\ln x$, the cases $B < 72$ yield poor results. For $x^y$, we tested with the $(x, y)$ domain specified in Section IV.B (this is not the full allowable domain); here, the cases $B < 40$ yield poor results.

For $e^x$ and $B = 24$, 16 integer bits is insufficient to properly represent many intermediate and output values, hence the poor accuracy results. This is illustrated in Fig. 10, where we plot the hardware outputs alongside the MATLAB® built-in function's. For $B = 24$, notice the point where the hardware starts producing incorrect values. For $B = 28$, the plots look identical, confirming that $IW = 20$ bits is the minimum required to properly represent the intermediate and output values.

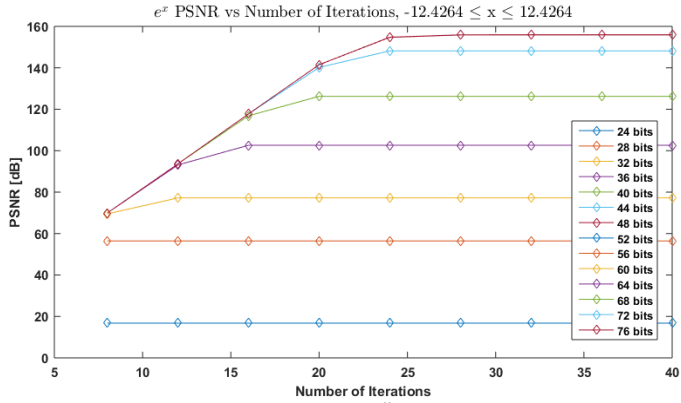

Figure 7. Accuracy (PSNR) results for $e^x$. Accuracy is high for $B > 24$.

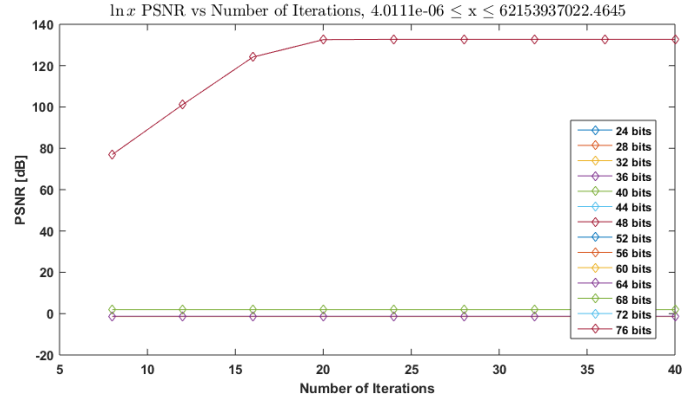

Figure 8. $\ln x$: Accuracy (PSNR) results. Poor accuracy occurs when we use fewer than 37 integer bits ($B < 72$) to represent the $\ln x$ domain ($M = 5$).

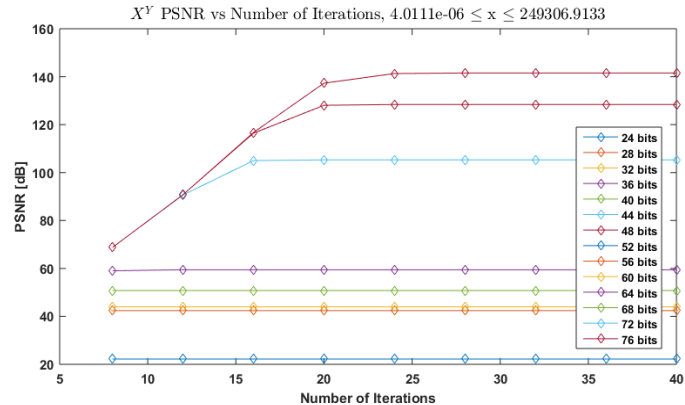

Figure 9. Accuracy (PSNR) results for $x^y$. Accuracy is high for $B > 24$.

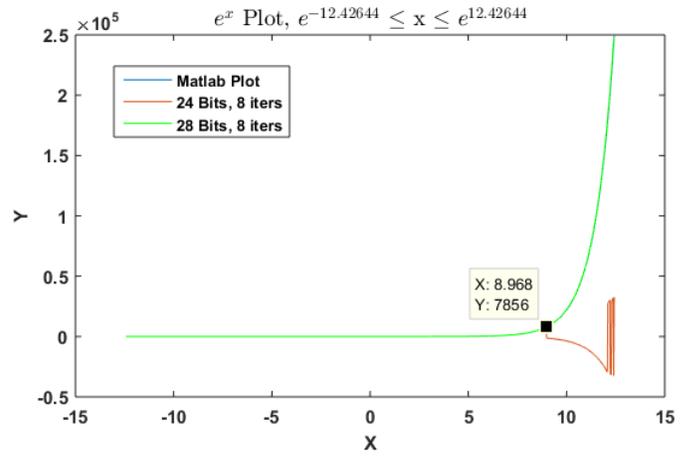

Figure 10. $e^x$: MATLAB® built-in function vs hardware results.

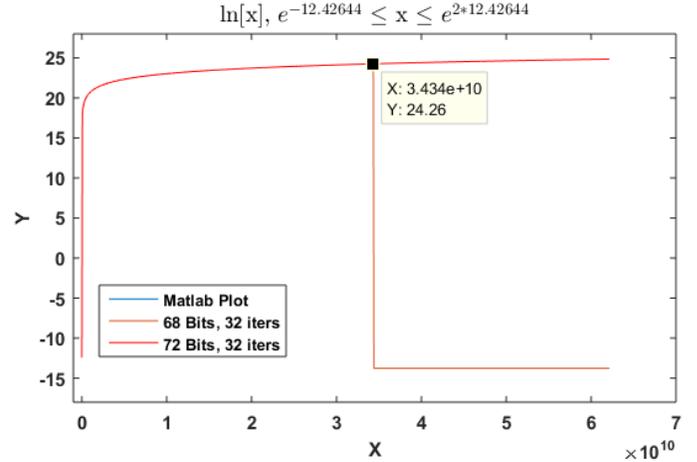

Figure 11. $\ln x$: MATLAB® built-in function vs hardware results.

For $\ln x$ (and thus $x^y$), note that if $B < 72$, then $IW < 37$. This does not properly represent the entire input domain of $\ln x$ (Table I for $M = 5$), hence the poor accuracy results. Fig. 11 illustrates this effect, where we plot our hardware's results versus the MATLAB® built-in function's. Note that for $B = 68$ ($IW = 36$), the maximum input value allowed by the [68 36] format is $3.44 \times 10^{10}$ (smaller than what Table I allows for $M = 5$). This is exactly the point at which the $\ln x$ curve starts producing incorrect values. Thus, for $B < 72$, $\ln x$ can only produce correct values if we restrict the input domain to what the fixed point format can represent.

As for $x^y$, we detailed some issues when using the full allowable convergence domain for $e^x$ and $\ln x$, this provides a hint on the behavior of $x^y$. Fig. 12 depicts the $x^y$ plot for $B = 28, 40$ and the domain: $x \in [e^{-\theta_{max}(M)/2}, e^{\theta_{max}(M)/2}], |y| \leq 2$, this 'box' allows for a good depiction of the $x^y$ surface. Notice how for $B = 28$, the $x^y$ plot differs significantly in some areas when compared to the relatively accurate case with $B = 40$.

### D. Multi-objective optimization of the design space for $x^y$

Since the execution time depends solely on the $M$ and $N$, we consider it more important to illustrate the trade-offs between accuracy and resources. We present the accuracy-resources plot for all design parameters for $x^y$ in Fig. 13. This allows for a rapid trade-off evaluation of resources (given in Zynq-7000 slices) and accuracy for every generated hardware profile.

Moreover, Fig. 13 also depicts the Pareto-optimal [10] set of architectures that we extracted from the design space. This allows us to discard, for example, hardware profiles ($B > 52$) that require more resources for no increment in accuracy. The figure also indicates the design parameters that generate each Pareto point. There are hardware realizations featuring poor accuracy (less than 40 dB) in the Pareto front. For design purposes, these points should not be considered.

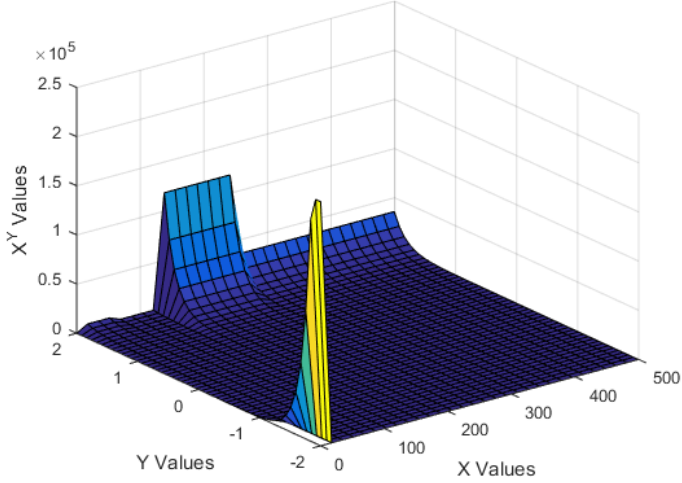 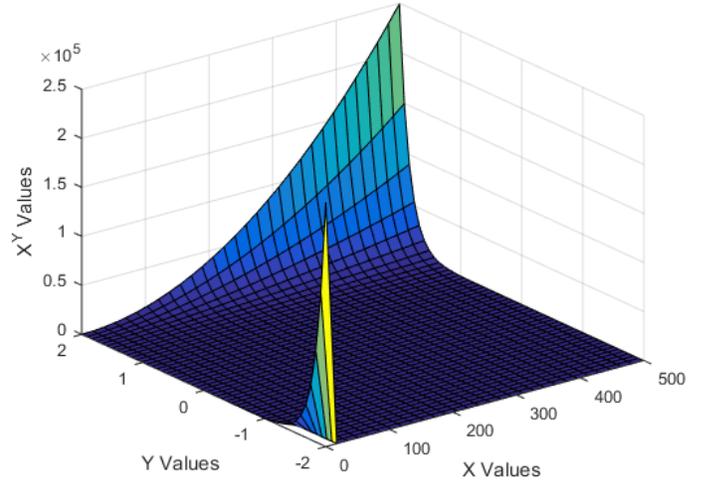

Figure 12. 3-D plot for $x^y$. $x \in [e^{-\theta_{max}(M=5)/2}, e^{\theta_{max}(M=5)/2}]$, $|y| \leq 2$ for format [28 8] and [40 20]. Note that inaccuracies when $B = 28$ (format [28 8])

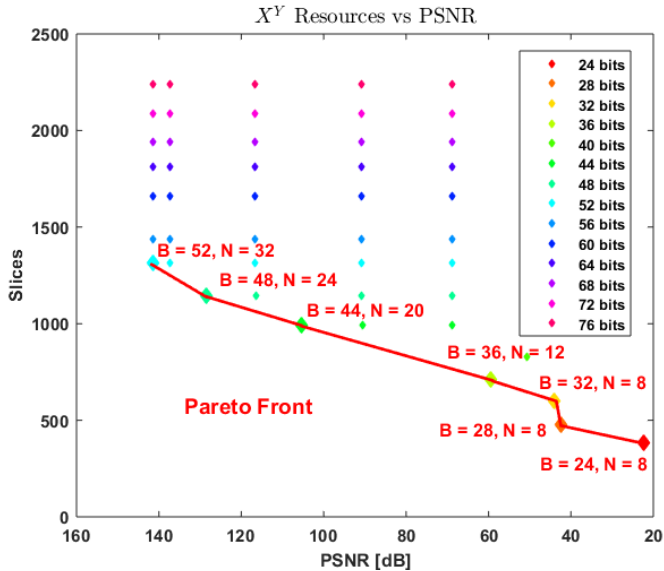

Figure 13. Resources-accuracy space and Pareto-optimal front for $x^y$. Device: Zynq-7000 XC7Z010 SoC with 4,400 Slices.

This approach allows the user to select only Pareto-optimal hardware realizations that simultaneously satisfy resources and/or accuracy constraints. For example: *i) highest accuracy regardless of resource usage*: the hardware with format [52 32] and $N = 32$ iterations satisfies this requirement at the expense of a large resource usage, *ii) lowest resource usage subject to accuracy $\geq 100$ dB*: the hardware with format [36 16] and $N = 12$ iterations satisfies the constraint and minimizes resource usage, *iii) lowest resource usage for accuracy $\geq 40$ dB*: the hardware with format [28 8] and $N = 8$ meets this constraint, *iv) highest accuracy for less than 1000 Slices*: the hardware with format [44 24] and $N = 20$ meets these constraints, and *v) Accuracy > 40 dB for no more than 1000 Slices*: Three hardware profiles satisfy these constraints. We select the one that further optimizes a particular need: accuracy or resources.

## VI. CONCLUSIONS

A fully parameterized fixed point iterative architecture for $x^y$ computation was presented and thoroughly validated. The expanded CORDIC approach allows for customized improved bounds on the domain of $x^y$. The Pareto-optimal architectures extracted from the multi-objective design space allows us to solve optimization problems subject to resources and accuracy constraints. We also provided a comprehensive assessment of how the fixed-point architecture affects the functions.

Further efforts will focus on the implementation of a family of architectures for $x^y$, ranging from the iterative version presented here to a fully pipelined version. We will also explore the use of the scale free hyperbolic CORDIC [5] which requires fewer iterations for the same interval of convergence.

REFERENCES

[1] J. Piñeiro, M.D. Ercegovac, J.D. Brugera, "Algorithm and Architecture for Logarithm, Exponential, and Powering Computation," *IEEE Transactions on Computers*, vol. 53, no. 9, pp. 1085-1096, Sept. 2004.

[2] F. De Dinechin, P. Echeverría, M. López-Vallejo, B. Pasca, "Floating-Point Exponentiation Units for Reconfigurable Computing," *ACM Trans. on Reconfigurable Technology and Systems*, vol. 6, no. 1, p. 4, May 2013.

[3] P. Meher, J. Valls, T.-B. Juang, K. Sridharan, K. Maharatna, "50 Years of CORDIC: Algorithms, Architectures, and Applications", *IEEE Trans. on Circuits and Systems I: Regular Papers*, vol. 56, no. 9, Sept. 2009.

[4] D. Llamocca, C. Agurto, "A Fixed-point implementation of the expanded hyperbolic CORDIC algorithm," *Latin American Applied Research*, vol. 37, no. 1, pp. 83-91, Jan. 2007.

[5] S. Aggarwal, P. Meher, K. Khare, "Scale-free hyperbolic CORDIC processor and its application to waveform generation," *IEEE Trans. on Circuits and Syst. I: Reg. Papers*, vol. 60, no. 2, pp. 314-326, Feb. 2013.

[6] J. C. Bajard, S. Kla, and J. Muller, "BKM: A new hardware algorithm for complex elementary functions," *IEEE Transactions on Computers*, vol. 43, no. 8, pp. 955-963, August 1994.

[7] V. Kantabutra, "On Hardware for Computing Exponential and Trigonometric Function," *IEEE Transactions on Computers*, vol. 45, no. 3, pp. 328-339, March 1996.

[8] X. Hu, R.G. Harber, S.C. Bass, "Expanding the range of convergence of the CORDIC algorithm," *IEEE Transactions on Computers*, vol. 40, no. 1, pp. 13-21, Jan. 1991.

[9] J. Mack, S. Bellestri, D. Llamocca, "Floating-Point CORDIC-based Architecture for Powering Computation," to appear in *Proceedings of the 10th International Conference on Reconfigurable Computing and FPGAs (ReConFig'2015)*, Mayan Riviera, Mexico, December 2015.

[10] S. Boyd and L. Vanderberghe, *Convex Optimization*. Cambridge, U.K.: Cambridge Univ. Press, 2004.